\documentclass{new_tlp}
\pdfoutput=1
\usepackage{xspace}
\usepackage{latexsym}
\usepackage{amssymb}   % for \mathbb

\usepackage{ifthen}
\newboolean{commentsaon}         % to control the top level of comments
\setboolean{commentsaon}{true}
  \setboolean{commentsaon}{false}  % to remove both kinds of comments

\newboolean{commentson} % to control comments
\setboolean{commentson}{true}
\newcommand{\comment}[1]
{\ifthenelse{\boolean{commentson}\AND\boolean{commentsaon}}
   {{\par\noindent\mbox{}{\small\footnotesize\blue[ *** #1 ]\par}\noindent\par}}
   {}}

\newcommand{\commenta}[1]
{\ifthenelse{\boolean{commentsaon}}
   {{\par\noindent\mbox{}{\small\color[rgb]{0, .5, 0}[ *** #1 ]\par}\noindent\par}}{}}

\usepackage{wrapfig}

\usepackage{pgf}

\usepackage[yyyymmdd]{datetime}

\usepackage{color}
\newcommand\blue     {\color{blue}}

\newcommand{\dblue}{\color[rgb]{0,0,.9}}
\newcommand{\dgrey}{\color[rgb]{.4, .4, .4}}

\usepackage{hyperref}

\newtheorem{example}{Example}
\newtheorem{definition}{Definition}

\newcommand*{\seq}[2][n]  {{#2_{1}, \allowbreak \ldots, \allowbreak #2_{#1}}}

\newcommand*{\mydash}{{\mbox{\tt-}}}
\newcommand*{\myprologneg}{\ensuremath{\mbox{\tt \symbol{92}+}}\xspace}
\newcommand*{\myunderscore}{\mbox{\tt\symbol{95}}}

\newcommand{\QQ}{{\ensuremath{\cal{Q}}}\xspace}
\newcommand{\DD}{{\ensuremath{\cal{D}}}\xspace}
\newcommand{\TT}{{\ensuremath{\cal{T}}}\xspace}

\title
     [Implementing backjumping]
    {Implementing backjumping by means of exception handling}
\author[W. Drabent]
{W{\l}odzimierz Drabent%
        {\ifthenelse{\boolean{commentson}\AND\boolean{commentsaon}}
           {\blue\quad  [with private comments]}{}%
        }%
        \\
        \small
      \begin{tabular}{c} \\[-1ex]
           Institute of Computer Science,
           Polish Academy of Sciences%
          \\ %
 \mbox{{\tt drabent\,{\it at}\/\,ipipan\,{\it dot}\/\,waw\,{\it dot}\/\,pl}}
      \end{tabular}
}

\submitted{\today} \revised{} \accepted{}  %
\begin{document}

\maketitle

\begin{abstract}
We discuss how to implement backjumping (or intelligent backtracking) in Prolog
by using the built-ins throw/1 and  catch/3.
We show that it is impossible in a general case, contrary to a claim that
``backjumping is exception handling".
We provide two solutions.   One works for binary programs; in a general case it
imposes a restriction on where backjumping may originate.
The other restricts the class of backjump targets.
We also discuss implementing backjumping by using backtracking and the Prolog
database. 
  Additionally, we explain the semantics of Prolog exception handling in
  the presence of coroutining.

\end{abstract}


\section{Introduction}
\label{sec.introduction}
In this note we first explain the incompatibility between 
backjumping (or intelligent backtracking) and the exception handling of Prolog.
We show that in general it is impossible to implement the former by the latter.
Then show how to do this for some restricted cases 
(Section \ref{sec.implementing}).
We present two approaches.
The first one is applicable 
to a restricted but broad class of cases,
including binary programs with arbitrary backjumping.  
For non-binary programs
the restriction constrains from where the backjumping may originate.
In the second approach the class of available backjump targets is restricted,
so the resulting backjumping may only be an approximation of that intended.
Section \ref{sec.examples} presents an example of each approach.
The next section discusses implementing backjumping by means of backtracking
and the Prolog database.
The report is concluded by a brief discussion of the related work and
conclusions (Section \ref{sec.final}).
In an appendix we explain the semantics of Prolog exception handling in a
presence of coroutining (also known as delays).
The main motivation for this work were opinions that exception handling
was an ideal tool to implement backjumping
\cite{DBLP:journals/tplp/RobbinsKH21}.

\paragraph{Preliminaries.}
This paper employs the standard terminology and basic
well-known notions of logic programming  \cite{apt-prolog}.
So ``atom'' means atomic formula, and the nodes of SLD-trees are
queries, i.e.\ sequences of atoms.
Unless stated otherwise, we consider LD-resolution, i.e.\
SLD-resolution under the Prolog selection rule.
So we do not deal with coroutining/delays.  
By a {\em $p$-atom} we mean an atom %
with the predicate symbol $p$.
{\em Procedure}~$p$ of a program $P$ is the set of the clauses of $P$ beginning
with $p$. 
By an {\em answer} (resp. {\em computed} answer) %
 of a program $P$ with a query $Q$
 we mean $Q\theta$ where $\theta$
 is a correct (computed) answer substitution for $P$ and $Q$.

%

%
%
%
%

%

% \smallskip
\medskip
 {\small
Now we formalize some concepts often seen as obvious.
The rest of this section may be skipped at the first reading,
as it is only needed for formal aspects of few fragments of this paper.

Popular ways of explaining Prolog, like the Byrd box model \cite{ByrdBox},
treat the atom selected in a query as a procedure call.
The issue of what is the execution of such atom is often
supposed to be obvious, and left unexplained. %
For instance the Prolog standard  \cite{Prolog.standard96}
uses a notion  ``the execution of'' but does not seem to define it
   (however some hint is given in  Fig.\,4.5). %
Here we provide such a definition 
in terms of LD-trees, following the idea of \citeN[Def.\,5.1,\,5.2]{DM88}.

Consider a (possibly infinite) LD-derivation $\DD=Q_0,Q_1,\ldots$ and a query
$Q_i=A,B$ from \DD (where $A,B$ are sequences of atoms, and $A$ is nonempty).
The {\em execution} in \DD of (the occurrence of) $A$ in $Q_i$ is the part
$\DD'$ of \DD consisting
of those queries $Q_k$ ($k\geq i$) which are of the form
$Q_k=A',B\theta$, and for $i\leq j<k$,
 each query $Q_j$ is not an instance of $B$.
 In other words, the execution $\DD'$ of $A$ consists 
of the queries of \DD starting from $Q_i=A,B$
 up to the first query of the form $B\theta$, if such query exists.
(So otherwise $\DD'$ contains
$Q_i$ and all the following queries.)
 If such query exists, the execution 
$\DD' = (A,B),\ldots,(B\theta)$  is called {\em successful}.
%%%
% Note that $A\theta$ is the computed answer for $A$.
We may assume that $\theta$ is the composition of the mgu's used in the
% resolution steps from $Q_i$ to $B\theta$.  
LD-derivation $\DD'$.  
Then $A\theta$ is the computed answer for $A$.

Now consider a query
 $Q_l=B',A,B$ (with nonempty $B'$ and $A$) in \DD, and assume that
\DD contains a query $Q_j$, $l<j$, which is an instance of $A,B$.
Let $Q_i=(A,B)\theta$ be the the first such query.
Then by the %
 {\em execution} in \DD of (the occurrence of) $A$ in $Q_l$ we mean
the execution in \DD of (the occurrence of) $A\theta$ in $Q_i$.
Let $Q=B',A,B$  (with nonempty $A$) be a node in an LD-tree \TT. 
Consider the derivations that are the branches of \TT containing $Q$.
The {\em execution} 
of the occurrence $A$ in $Q$ in \TT (briefly, the execution of $A$)
is the subgraph of \TT consisting of the executions of $A$ in $Q$ in these
derivations.
Note that if $B'$ is empty then the execution is a tree, otherwise it is a
forest. 

   } %

\section{Backjumping and Prolog exception handling}
\label{sec.backjumping}
In this section we present and compare backjumping and the Prolog exception
handling. 
\subsection{Backjumping}
A Prolog computation can be seen as a depth-first left-to-right traversal of an
SLD-tree.
Each node with $i$ children is visited $i+1$ times
(it is first entered from its parent and then from each of its children).  
Moving from a node to its
parent is called {\em backtracking}.  By {\em backjumping} we mean skipping a part of
the traversal, by moving immediately from a node to one of its non-immediate
ancestors (called the backjumping {\em target}\/).
{\em Intelligent backtracking} \cite{BruynoogheP84}
is backjumping in which it is known that there are no successes in the
omitted part of the SLD-tree.

 Let us a have a more detailed look.
Consider a node $Q$ with $k$ children $Q_1,\ldots,Q_k$ (Fig.\,\ref{figureQ}).
We may say that $k$ backtrack points correspond to $Q$; 
backtracking from $Q_i$ means arriving to $Q$ at its $i$-th backtrack point.
Obviously,
this is followed by visiting $Q_{i+1}$ when $i<k$,
and by backtracking to the parent  of $Q$ when $i=k$.
Similarly, backjumping to $Q$ from a node in the subtree
rooted in $Q_i$ should arrive at the $i$-th backtrack point.
Thus such backjumping is followed by visiting $Q_{i+1}$ when $i<k$,
and the parent of $Q$ when $i=k$.

One may consider a generalization of backjumping, where not only a part of
the subtree  rooted in $Q_i$ is skipped, but also the subtrees rooted in 
$Q_{i+1},\ldots,Q_j$ (for some $j\in\{i{+}1,\ldots,k\}$).
We do not discuss such a generalization here.

\begin{figure} 
\[
\begin{array}{c}
\dgrey N \\
 | \\[.3ex]
Q  \\
\newcommand{\wys}{2.5ex}
\newcommand{\szer}{12ex}
\newcommand{\szera}{2.8ex}  %
{%
    \begin{pgfpicture}
	  \pgfpathmoveto{\pgfpoint{-2ex}{0ex}}
	  \pgfpathlineto{\pgfpoint{-\szer}{-\wys}}
	  \pgfpathmoveto{\pgfpoint{2ex}{0ex}}
	  \pgfpathlineto{\pgfpoint{\szer}{-\wys}}

	  \pgfpathmoveto{\pgfpoint{-1ex}{-0.5ex}}
	  \pgfpathlineto{\pgfpoint{-\szera}{-\wys}}
	  \pgfpathmoveto{\pgfpoint{.3ex}{-0.5ex}}
	  \pgfpathlineto{\pgfpoint{\szera-1.2ex}{-\wys}}
	  \pgfusepath{stroke}
    \end{pgfpicture}%
}
\\[-.5ex]
\begin{array}[t]{cccccc}
Q_1 & \cdots & Q_i & Q_{i+1}\hspace*{-.4em} & \cdots\hspace*{-.2em} & Q_k
\\
&&
  \newcommand{\mysize}{10ex}
  \makebox[0pt]
          {%
      \begin{pgfpicture}
            \pgfpathmoveto{\pgfpoint{-1ex}{\mysize}}  %
            \pgfpathlineto{\pgfpoint{-\mysize}{0ex}}
            \pgfpathlineto{\pgfpoint{\mysize}{0ex}}
            \pgfpathlineto{\pgfpoint{1ex}{\mysize}}
            \pgfusepath{stroke}
            \pgfputat{\pgfpoint{0pt}{3ex}}{\makebox[0pt]{${\cal S}$}}
      \end{pgfpicture}%
          }\,\,
\end{array}
\end{array}
\]  
\label{label.proba}%
  \caption{%
 \label{figureQ}
\small
 Backjumping to $Q$ from the subtree ${\cal S}$ rooted in $Q_i$ (when $i<k$)
  is followed by visiting $Q_{i+1}$.
  For $i=k$ the backjumping is followed by visiting the parent of $Q$.
\\
\mbox\quad
    Consider exception handling 
    (implemented by making $N$ to be $catch(Q,s,H)$).
    Exception $s$ raised in ${\cal S}$ (and not caught in ${\cal S}$) 
    results in skipping $Q_{i+1},\ldots,Q_k$ and all their descendants.
    Additionally, the LD-tree is extended by a subtree corresponding to
    executing the exception handler $H$.  %
\\[-.5ex]\mbox{}\hrulefill
}
\end{figure}

\subsection{Exception handling}
\label{sec.exceptions}
Prolog provides an exception handling mechanism, consisting of built-in
predicates {\it throw}/1 and {\it catch}/3.
Let us follow the Prolog standard
and explain them in terms of LD-trees.
 Assume that a ${\it catch}$-atom $A_c={\it catch(Q,s,Handler)}$
is selected in a node $N$ of an LD-tree.  
This means the node is of the form
$N = A_c,N'$.  It is required that $Q$ and ${\it Handler}$ are queries.
Informally, the execution of $A_c$ means executing $Q$. 
More precisely, node $A_c,N'$ has a single child $Q,N'$.  A second child
may however be created as a result of exception handling.

An exception $t$ is raised by
invoking ${\it throw(t)}$.  Formally, the tree has a node
 $N_t={\it throw(t),N_t'}$; such node has no children.
This is sometimes called {\em throwing a ball $t$}.
Visiting $N_t$ starts a search along the
path from $N_t$ to the root.  The search is for a node
$N_c={\it catch(Q',s',Handler')},N_c'$ with one child
 such that

\smallskip
 \quad
\begin{oldtabular}{l}
   (a) a freshly renamed copy $t'$
   of the ball $t$ is unifiable with $s'$, with an mgu $\theta$, and
   \\
   (b$'$) ``the ball is thrown during the execution of'' $Q'$
   \cite{Prolog.standard96}.    
 \end{oldtabular}
\\[\smallskipamount]
It is useful to provide a more formal wording for the latter
%%%
(cf.\ Preliminaries, Section \ref{sec.introduction}):
% (employing a definition from Section \ref{sec.introduction}):
  %

\smallskip
\quad
 \begin{oldtabular}{l}
   (b)
   \parbox[t]{.89\textwidth}
   {
   no node between $N_c$ and $N_t$ (including  $N_t$)  is an instance of $N_c'$
   (this means that each of these nodes is  of the form  $M,N_c'\sigma$,
   with a nonempty $M$).
   }
 \end{oldtabular}
\\[\smallskipamount]
The first (closest to $N_t$) such node $N_c$ on the path is chosen, and
 a new child ${\it (Handler',N_c')\theta}$ of $N_c$ is added to the tree.
The new child becomes the next visited node of the tree.
(It is an error if such node $N_c$ is not found.)
{\sloppy\par}

\subsection{Implementing backjumping}
\label{sec.backjumping-exceptions}
Prolog does not provide any way to directly implement backjumping.
It may seem that exception handling
is a suitable tool for this task.  There is however an important difference.
Backjumping to a query $Q$ cannot be implemented by $catch(Q,s,H)$,
as this results in arriving at the last backtrack point corresponding to~$Q$
(cf. Fig.\,\ref{figureQ}).
All the unexplored descendants of $Q$ are omitted.
(The same happens if
we instead use $catch(Q',s,H),Q''$, where $Q=Q',Q''$.
Also, the omitted part of the tree cannot be explored by reconstructing it
by the exception handler $H$,
at least in the general case.
Roughly speaking, 
the handler may re-execute $Q$, but it does not have access to information
making it possible to skip the already visited children of $Q$.

This shows that backjumping to a node $Q$ cannot be implemented by
augmenting $Q$ with $catch$.  In the next section we show how to obtain such
backjumping by adding $catch$ (with ${\it fail}$ as the handler) to each
child of $Q$.

The scope of  ${\it catch}$/3 is the reason of the main limitation;
 ${\it catch(Q,\ldots)}$ cannot
intercept an exception raised outside the execution of $Q$.
Consider a query %
$A,B$ and assume that backjumping is to be
performed from the execution of $B$ %
to the execution of $A$. %
To implement it by means of exception handling, {\it catch} %
 has to appear in a
clause used in the execution of $A$.
But then it cannot catch an exception raised in the execution of $B$.

This discussion shows that backjumping cannot be, in general, directly
implemented by means of Prolog exception handling.

  We should also mention differences not related to implementing backjumping.
  In exception handling,  after an exception is caught, the exception
    handler
  is activated.  So the search space is augmented.
  In backjumping
    there is nothing similar to an exception handler; the search space is not
    modified. 
  Also, in contrast to backjumping, exception handling makes it possible to
  pass information (an arbitrary term) from the point where the exception is
  raised to the one where it is caught.  This is done by means of the
  argument of  ${\it throw}$/1.

\section{Employing exception handling}
\label{sec.implementing}
This section introduces two restricted approaches to implement backjumping.
Examples 
are given in the next section.
We consider Prolog without coroutining/delays. 

\subsection{Approach 1}
\label{sec.approach1}
Assume that we deal with a definite clause program $P$, which we want to
execute with backjumping.
The target of backjumping is to be identified by a term $id$.  
So backjumping is initiated by $throw(id)$.

Assume that the
target of backjumping is a node $A,Q$ of the LD-tree, where $A$ is an atom.
Assume that $A,Q$ has $k$ children, $\seq[k]Q$.
Let $p$ be the predicate symbol of $A$ and 
\begin{equation}
\label{theprogram}
\begin{array}{l}
  p(\vec t_1)\gets B_1. \\ \cdots \\     p(\vec t_n)\gets B_n.
\end{array}
\end{equation}
where $k\leq n$,
be the procedure $p$ of program $P$
(i.e.\ the clauses of $P$ beginning with $p$).

Consider backjumping initiated by $throw(id)$ in the subtree rooted in $Q_i$.  
The subtree should be abandoned, but the descendants of $Q_{i+1},\ldots,Q_k$
should not. Thus we need to restrict the exception handling to this subtree.
A way to do this is to replace each $B_j$ by
$
catch( B_j, id, {\it fail} )
$.
Then performing $throw(id)$ while executing $B_j$ results in failure of the 
clause body and backtracking to the next child of $A,Q$, as required.
Assume that a query $b t id(\vec t,Id)$ 
\newcommand{\myunderline}[1]{\makebox[0pt][l]{\underline{#1}}}%
(\myunderline{b}backjump \myunderline{t}target \myunderline{id}identifier)
 produces out of the arguments $\vec t$ of $p$ the unique identifier %
of $A,Q$ as the backjump target.
Now,  the backjumping is implemented by a transformed procedure consisting of
clauses
\begin{equation}
\label{eq.approach1.1}
{\it
p(\vec t_j)\gets b t id(\vec t_j,Id), \,
  catch( B_j, Id,  fail ) }.
\qquad\qquad \mbox{for }j=1,\ldots,n
\end{equation}
(where ${\it Id}$ is a variable).

Transforming a program in this way correctly implements backjumping, however
with an important limitation.
Backjumping to a node
$p(\vec t),Q$ must occur during the execution of
$p(\vec t)$.
Otherwise the exception is not caught and the whole computation terminates
abnormally.

An important class of programs which satisfy this limitation are binary logic
programs (i.e.\ programs with at most one body atom in a clause).  
The approach presented here works for
such programs and arbitrary backjumping.
It also works when each backjump target is a node consisting of a single atom.

Sometimes (like in Ex.\,\ref{ex.binary} below)
it may be determined in advance that, for some $j$, no exception
will be caught by the ${\it catch}$/3 in  (\ref{eq.approach1.1}).
So in practice
some clauses of (\ref{theprogram}) may remain unchanged
(or a choice between $B_j$ and  ${\it catch( B_j, Id,  fail)}$
may be made dynamically,
e.g.\ 
 by modifying the body of  (\ref{eq.approach1.1}) into
${\it
b t id(\vec t_j,Id) %
    \to    %
      catch( B_j, Id, {\it fail} )  %
    \linebreak[3]
\mathop;
 B_j 
}
$).

\paragraph{Approach 1a.}
\label{proba.paragraph}
Here we present a variant of Approach 1.
Roughly speaking, in the former approach
control is transferred to the next clause due to failure of a clause body.
So catching an exception causes an explicit failure.
Here control is transferred to the next clause by means of an exception
% , so standard backtracking eventually raises an exception.
% (and standard backtracking is implemented by means of exceptions).
(hence standard backtracking is to be implemented by means of exceptions).
To simplify the presentation we assume that in  (\ref{theprogram}) all
the clause heads are the same, 
$\vec t_1=\cdots=\vec t_n=\vec t$.

Assume first that $n=2$.  Backjumping equivalent to that of Approach 1 can be
implemented by
%%%%
\vspace{-2ex plus 1ex}
% \vspace{-2ex plus 2ex}
%
\begin{equation}
\label{program.approach1asmall}
\begin{array}{l}
\it
p(\vec t) \gets  b t id(\vec t,Id),
     catch(
     \begin{array}[t]{l}
\it      (B_1 \mathrel; throw(Id)),   \\     
\it       Id,  \\
{\it       catch( B_2, Id, {\it fail})\ ).}
     \end{array}
\end{array}
%%%%
\vspace{1ex minus 1ex}
\end{equation}
Invocation of $B_2$ is placed in the exception handler,
so we additionally raise an exception when $B_1$ (the first clause body) fails.
For arbitrary $n$, the transformed procedure (\ref{theprogram}) is:
%%%%
%\vspace{-2ex plus 1 ex}
\begin{equation}
\label{program.approach1a}
\begin{minipage}{.9\textwidth}
%%%%
\vspace*{-1\abovedisplayskip plus 1ex minus 1ex}
 \[
    p(\vec t) \gets
    \begin{array}[t]{l}
\it     b t id(\vec t,Id), \\
         catch(
         \begin{array}[t]{l} (\it B_1 \mathrel; throw(Id)), \\
\it                    Id ,              \\
         catch(
         \begin{array}[t]{l} (B_2 \mathrel; throw(Id)), \\
\it                    Id, \\ \ldots \\
         catch(
         \begin{array}[t]{l} (B_{n-1} \mathrel; throw(Id)), \\
\it                    Id ,              \\
\it         catch(
                    B_{n},\,  %
                    Id ,\,             %
                    {\it fail} \,)\ \, )\cdots )).
    \end{array}\end{array}\end{array}\end{array}
    \hspace*{-3em}
\]
\end{minipage}%
\end{equation}

Generalizing this transformation to clauses with different heads
is rather obvious. 
 Note that 
in this approach it is possible to augment backjumping by passing information
(from the place where the backjump originates to the backjump target).%
%\ Such augmenting is impossible in Approach 1 and Approach 2 below.
%a
% \smallskip
% \vspace{.5ex}
 \footnote
{%\small
To pass a term $t$, one may choose the backjump target identifier to be 
$f(X_i)$ for clause $i$.  Then performing ${\it throw}(f(t))$ while executing
$B_i$ results in binding $X_i$ to $t$ when the exception is caught.
This makes $t$ available in those bodies $B_{i+1},\ldots,B_n$ that contain
$X_i$.  
E.g.\ for $n=2$
instead of the body of (\ref{program.approach1asmall}) we obtain
${\it
catch(\, (B_1 ; throw(f(no b j))),\, f(X_1),\, 
    catch(B_2,   f(X_2), fail) \, )
}$;  constant ${\it no b j}$ (for ``no backjumping'') is passed when 
standard backtracking takes place.
%
% \vspace{1.5ex}  % {.5ex}
} % \footnote
 Such augmenting is impossible in Approach 1 and Approach 2 below.

\subsection{Approach 2, approximate backjumping}

\newcommand\stemsa{%
\raisebox{2ex}[0pt][0pt]{%
    \makebox[0pt]{%
    \begin{pgfpicture}
    \pgfsetdash{{0.5pt}{1.5pt}}{0pt}

    \pgfqmoveto{-2}{1}
    \pgfqcurveto{-54}{-8}{-42}{-22}{-25}{-36}    %

     \pgfqmoveto{8}{0}
     \pgfqcurveto{70}{-7}{45}{-35}{7}{-61}       %

    \pgfqmoveto{-5}{-18} 
    \pgfqcurveto{-5}{-26}{-11}{-32}{-18}{-36}      %

    \pgfqmoveto{11}{-18} 
    \pgfqcurveto{18}{-34}{19}{-48}{5}{-61}      %

    \pgfqmoveto{7}{-45} 
    \pgfqcurveto{8}{-50}{8}{-55}{3.5}{-61}      %
    \pgfusepath{stroke}
    \end{pgfpicture}
}}%
}

\newcommand{\myfigBBcorr}{%
% \makebox[0pt][l]{................}
  \begin{minipage}[t]
 {.287\textwidth}
% {.26\textwidth}
% {.285\textwidth}  % for without Fig
\vspace*{-4ex}
% {-6ex}  % for a wider wrapfig
\[
      \begin{array}[t]{c}
        N_0=A_0,Q_0\ \  \\
 \makebox[0pt][l]{\stemsa}\hspace{.5pt}
 \raisebox{-.4ex}
          {\rule{.4pt}{2ex}} \\
        N_1=(B_0,B_1,Q_0)\theta\\[-.5ex] \vdots \\[.5ex]
        {\dblue N}=  A,Q \qquad \   %
        \\[-.5ex] \vdots \\[.5ex] 
       {\dblue N'}
   %
%         \\ [2ex]
     \end{array}
\]
\footnotesize
Fig.\,2.
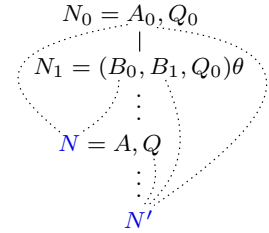
    Dotted lines connect an atom sequence with a query from its execution.
% \vspace{1ex}
%
    $N,N'$ are of the form
\vspace{1.3ex}

%\hspace{.5em}%
\mbox{}\hfill
    \begin{oldtabular}{@{}l@{}}
      $N=A,\ldots,(B_1,Q_0)\theta\rho$, \\            
      $ N'={\it throw(id)},\ldots,Q_0\theta\rho\varphi$.
    \end{oldtabular}%\hspace*{-1em}
\vspace*{-1ex}
    \end{minipage}%
}  % \myfigBBcorr

\noindent
We have shown how to implement backjumping to an LD-tree node $N = A,Q$
(with atomic $A$)
from within the execution of $A$.
It remains to discuss backjumping originating in the execution of $Q$.

Assume that the initial query is atomic; dealing with arbitrary initial
queries is similar.
In such case, the program contains a clause $H{\gets}B_0,B_1$ %
(where $B_0,\,B_1$ are nonempty), %
such that, speaking informally,
 the backjumping is from within the execution of $B_1$,
and its target $N$ is within the execution of $B_0$. % \footnote was here

% \footnote{\label{bigfootnote}%
 \smallskip
%   \medskip
{\small
\begin{wrapfigure}{r}{.34\textwidth}
% {.29\textwidth}
  \footnotesize\hfill
              \raisebox{.9ex}[0pt]
              {\scalebox{1.1}{\myfigBBcorr}%
              }
%              \raisebox{2.8ex}[0pt][36ex]{\myfigBBcorr}
%            \myfigBBcorr%
\end{wrapfigure}
  Let us prove that such clause exists.
  Assume that the backjumping target is $N=A,Q$ ($A$ is an atom), and 
  that the backjumping originates in a node $N'$ in the execution of $Q$ in $N$.
  Let
  \DD be a branch of the LD-tree containing $N$ and $N'$.

  Obviously $N$ occurs in the execution in \DD of $A$ in $N$, and $N'$ does not.
  As the root of the tree is atomic, there exist in \DD 
  a node $N_0$ and its child $N_1$ such that $N$ and $N'$
  (i) occur in the execution in \DD of a single atom $A_0$ in $N_0$
  and (ii)
  do not occur in the execution in \DD of a single atom in $N_1$.
  Note that $A_0$ is the first atom of $N_0$ (otherwise (ii) does not hold).
  The two nodes are of the form $N_0=A_0,Q_0$ and $N_1=(Q',Q_0)\theta$,
  and a clause $H\gets Q'$ was used to obtain  $N_1$.

  Now $Q'$ can be split as $Q'=B_0,B_1$ (hence $N_1=(B_0,B_1,Q)\theta$),
  so that $N$ occurs in the execution of $B_0\theta$ in $N_1$,
  and $N'$ occurs in the execution of $B_1\theta$ in $N_1$.  
  So $H\gets B_0,B_1$ has the required property.

We are going to implement backjumping from $N'$ to the last node $N''$ of the
execution of $B_0\theta$ in $N_1$. 
To make  $N''$ as close as possible to $N$,
one should choose $B_0,B_1$ so that $N$ is in the execution of the last
atom of $B_0\theta$.

  % \vspace{1.5ex}
  % } % big \footnote
  %
}  % end \small

% \medskip
\smallskip

As discussed in Section  \ref{sec.backjumping-exceptions},
such backjumping
(from within the execution of $B_1$ to a target in the execution of $B_0$)
 cannot be implemented by means of {\it throw}/1 and 
{\it catch}/3.
What can be done is to force $B_1$ to fail when an
exception is thrown.
This means, speaking informally,  backjumping to the
success of $B_0$, instead of the original target $N$.
This in a sense approximates backjumping to $N$.  
In some cases such shorter backjumping may still be useful.
It may exclude from the search space a major part of what would
be excluded by backjumping to $N$.

The success of $B_0$ (formally of $B_0\theta$) is
    the topmost descendant of $N$ of the form  $N''=(B_1,Q_0)\theta\rho\psi$
    (here
     $\theta,\rho$, $Q_0,N',N''$ are as in Fig.\,2 and 
    the proof below).
%  footnote \ref{bigfootnote}).
Node $N''$ appears in the tree branch between $N$ and node $N'$
that originates the backjump.
To catch the exception at $N''$ we modify the program, 
so that the instance $B_1\theta\rho\psi$ of $B_1$
in node $N''$ is replaced by  $catch(B_1\theta\rho\psi,id,{\it fail})$.
% \footnote{%
% Backjumping implemented in this may be understood as arriving to the last
% backtrack point of $N''$, or -- due to the handler {\it fail} -- to 
% the parent of $N''$.
% }
\linebreak[3]
(The resulting backjumping may be understood as arriving to the last
backtrack point of $N''$, or -- due to the handler {\it fail} -- to 
the parent of $N''$.)
To obtain this, the clause
\[
H\gets B_0,B_1
\ \quad \mbox{ is transformed to } \ \quad
    H\gets B_0,\, b t id(\ldots,Id), \,
    catch(B_1,Id,{\it fail})
\]
where $b t id$, as previously, is used to obtain the unique identifier for
the backjump target.

\newcommand\tttrue{\mbox{\tt true-}}%
\newcommand\ttfalse{\mbox{\tt false-}}%

\enlargethispage{.3ex}  % to avoid a one line paragraph at the bottom
\section{Examples}
\label{sec.examples}
We apply the approaches introduced above to a simple program,
a naive SAT solver.  It uses the representation
of clauses proposed by  \citeN{howe.king.tcs-shorter}.
(Note that we deal here with two kinds of clauses -- those of the program, and
the propositional clauses of a SAT problem.)
A conjunction of clauses 
is represented as a list of (the representations of) clauses.
A clause is represented as a list of (the representations of) literals.
A positive literal is represented as a pair
$\tttrue X$ and a negative one as $\ttfalse X$, where the Prolog variable
represents a propositional variable.
For instance a formula
$(x\lor\neg y\lor z)\land(\neg x\lor v)$ is represented as 
{\tt[[true-X,false-Y,true-Z],[false-X,true-V]]}.
In what follows we do not distinguish literals, clauses, etc from their
   representations.

Thus solving a SAT problem for a conjunction of clauses $sat$
means instantiating the variables of $sat$ in such way that each of the lists 
contains an element of the form $t\mydash t$.
This can be done by a program $P_1$:  %
\[
   \begin{minipage}{.78\textwidth}  %
\begin{verbatim}
sat_cl( [Pol-Pol|Pairs] ).
sat_cl( [H|Pairs] ) :- sat_cl( Pairs ).
sat_cnf( [] ).
sat_cnf( [Clause|Clauses] ) :- 
                    sat_cl( Clause ), sat_cnf( Clauses ).
\end{verbatim}
   \end{minipage}
\]
and a query     ${\it sat\_c n f}(sat)$.
See %
the paper by \citeN[Section 3]{Drabent.tplp18} 
for further discussion and
a formal treatment of the program.

The examples below add backjumping to program  $P_1$.
They are not intended to provide a correct SAT-solver,
their role is only to illustrate Approaches 2 and 1.
In the examples, the backjumping is performed
after a failure of ${\it sat\_c l}(cl)$ (where $cl$ is the representation of a
partly instantiated clause).
The intended backjumping target is 
the last point where a variable from clause $cl$ was assigned a value.%

% \vspace{.5ex}
 \smallskip
% \footnote
{%
\small
  Note that such backjumping does not 
  correctly implement intelligent backtracking, some answers are lost.
  E.g. for $(x\lor y)\land(\neg z\lor z)\land(\neg x\lor\neg y)\land
  (\neg x\lor y\lor z)$ no solution with $z$ being \mbox{\tt true} is found.
  An explanation is that, speaking informally, backjumping from the last clause
  (with $x,y,z$ instantiated to
  {\tt true}, {\tt false}, {\tt false}
  or to
  {\tt true}, {\tt false}, {\tt false}) 
  arrives to the previous one
  (where $y$ or $x$ was set), this causes backjumping to the
  first clause.
\commenta{%
  An explanation is that, speaking informally, backjumping from the last clause
  (with $x,y,z$ instantiated to
  {\tt true}, {\tt false}, {\tt false})
  arrives to the previous one
  (where $y$ was set to false), this immediately causes backjumping to the
  first clause.
  }%  \commenta

}  % former \footnote

\commenta{
  Below we do not distinguish literals, clauses, etc from their
   representations.
}
\comment{Where?
   From now on \hspace*{-.3em}%
\hfil
  From v2, a{}r{}xiv: (?)
}

\begin{example}
\label{ex.approach2}
  Here we employ Approach 2 to program $P_1$.  Speaking informally, the
  required backjumping originates from within
 ${\it sat\_c n f(Clauses)}$  in the last clause of the program,
and its target is in     ${\it sat\_cl(Clause)}$.
 We approximate this backjumping by a failure of  ${\it sat\_c n f}$.
    (Note that in this case the approximation is good, 
    the intended target is a node of the form
     ${\it sat\_cl}([v\mydash V|t]),{\it sat\_c n f}(t')$
    and we implement backjumping to its child 
     ${\it sat\_c n f}(t'\{V/v\})$.)
\

We augment the values of variables; %
   the value of a variable is going to be of the form
 $(l,v)$, where $l$ is a number (the {\em level} of the variable)
   and $ v$ a logical value {\tt true} or {\tt false}.
The level shows at which recursion depth of ${\it sat\_c n f}$ the value was
assigned.  The levels will be used as identifiers for backjump targets. 
In such setting,
a substitution $\theta$ assigning values to variables makes a SAT
 problem $sat$ 
 satisfied when each member of list
 $sat\theta$ contains a pair
 of the form $ v\mydash(l, v)$.
 This leads to transforming the first clause of $P_1$ to 
 ${\it sat\_cl([Pol\mydash (\myunderscore,{\it Pol}) | {\it Pairs}])}$.

We transform $P_1$ into a program $P_2$ which takes levels into account.
We add the current level as the second argument
of ${\it sat\_c n f}$ and of ${\it sat\_c l}$.
A third argument is added to  ${\it sat\_c l}$;
it is used in finding the highest variable level in a clause.
The declarative semantics of the new program is similar to that of $P_1$;
the answers of $P_2$ are as follows.
  If
  the first argument of ${\it sat\_c l}$ is a list then it 
  has a member of the form $t\mydash(t',t)$.  
  Also, this condition is satisfied by each element of the list that is 
  the first argument of ${\it sat\_c n f}$.

  Operationally, an invariant will be maintained that, whenever 
  ${\it sat\_cl(cl,l,h l)}$ is selected in LD-resolution, $cl$ is a list and
  $l$ and $h l$ are numbers, 
  $l>h l$ and $l$ is greater than any number occurring in $cl$.
  List $cl$ is the not yet processed fragment of a clause $cl_0$
(possibly instantiated),
$l$ is the current level, and
${\it h l}$ is the highest level of those variables that occur in the already
processed part of $cl_0$ and
have been bound to some values at previous levels;
${\it h l}=-1$ when there is no such variable.
  In case of failure of ${\it sat\_c l(cl,l,h l)}$,
  an exception will be raised with the ball being the maximum of $h l$
  and the levels of the variables occurring in $cl$ (provided the maximum is
  $\geq0$).

Checking the value already assigned to a variable must be treated differently
from assigning a value to an unbound variable
(as its level is set only in the latter case).
This leads to two clauses playing the role of the first clause of $P_1$.
So procedure  ${\it sat\_c l}$ of  $P_1$ is transformed into the following
procedure of $P_2$:
%%%%
\vspace{0pt minus 0.6ex}
\[
\begin{minipage}[t]{.75\textwidth}
\begin{verbatim}
sat_cl( [Pol-V|_Pairs], _L, _HL ) :-
        nonvar(V), V=(_,Pol).
sat_cl( [Pol-V|_Pairs], L, _HL ) :-
        var(V), V=(L,Pol).
sat_cl( [_-V|Pairs], L, HL ) :-
        new_highest( V, HL, HLnew ),
        sat_cl( Pairs, L, HLnew ). 
\end{verbatim}
\end{minipage}
\makebox[0pt][l]{\hspace{3.3em}
    \begin{minipage}[t]{.05\textwidth}
        (\refstepcounter{equation}\theequation\label{clause1.ex.P2})
\\\\        (\refstepcounter{equation}\theequation)
    \label{clause2.ex.P2}
\\\\\\        (\refstepcounter{equation}\theequation)
    \label{clause3.ex.P2}
    \end{minipage}
} %
\]
Predicate ${\it new\_highest}$ takes care of updating the highest level
of the variables from the already processed part of the clause.
\[
   \begin{minipage}[b]{.76\textwidth}  %
{\small%
%%%%
  \begin{oldtabular}{@{}l@{}l}
    \% {\tt  new\_highest}${\it(var, h, h new)}$ -- \ & if
       ${\it var}$ is a Prolog variable
    then $h =h new$ \\
    \%     & otherwise ${\it var=(l, v)}$ and $h new =\max(h,l)$
    \vspace{-1ex}
  \end{oldtabular}
}
\begin{verbatim}
new_highest( V, H, H ) :- var( V ).
new_highest( V, H, H ) :- nonvar( V ), V=(L,_Value), H>=L.
new_highest( V, H, L ) :- nonvar( V ), V=(L,_Value), H<L.
\end{verbatim}
  \end{minipage}
\makebox[0pt][l]{\hspace{2.6em}%
    \begin{minipage}[b]{.05\textwidth}  %
\hfill   %
 (\refstepcounter{equation}\theequation\label{clause1aux.ex.P2})
\\ \mbox{}\hfill  
 (\refstepcounter{equation}\theequation\label{clause2aux.ex.P2})
\\ \mbox{}\hfill  
 (\refstepcounter{equation}\theequation\label{clause3aux.ex.P2})
    \end{minipage}
}  %
\]
Procedure ${\it sat\_c n f}$ is transformed into
%%%%
% \vspace{-.9ex}
%
\[
\begin{minipage}[t]{.46\textwidth}
\begin{verbatim}
sat_cnf( [], _L ).
sat_cnf( [Clause|Clauses], L ) :-
        sat_cl( Clause, L, -1 ),
        Lnew is L+1,
        sat_cnf( Clauses, Lnew ).
\end{verbatim}
\end{minipage}
\makebox[0pt][l]{\hspace{8.6em}%
    \begin{minipage}[t]{.05\textwidth}
\hfill   %
 (\refstepcounter{equation}\theequation\label{clause7.ex.P2})
\\\\[1.5ex] \mbox{}\hfill  
 (\refstepcounter{equation}\theequation\label{clause8.ex.P2})
    \end{minipage}
}  %
\]
\newcommand*{\myindent}{\hspace{5em}}%
Program $P_2$ consists of clauses (\ref{clause1.ex.P2}) -- (\ref{clause8.ex.P2}).
An initial query
${\it sat\_c n f}(sat,0)$ results in checking the
satisfiability of a conjunction of clauses $sat$.
Now we add backjumping to $P_2$.
The backjumping has to be triggered instead of a failure of ${\it sat\_cl}$.
The latter happens when the first argument of  ${\it sat\_cl}$ is $[\,]$.
The new program $P_3$ contains the procedure ${\it sat\_cl}$ of $P_2$, and
additionally a clause
\begin{equation}
\label{clause.throw.ex.P3}
\begin{minipage}[t]{.74\textwidth}
\begin{verbatim}
sat_cl( [], _, HL ) :- HL>=0, throw( HL ).
\end{verbatim}
\end{minipage}
\end{equation}
triggering a backjump.
When ${\it H L}<0$  then there is no target for backjumping, 
and standard backtracking is performed.

The procedure ${\it sat\_c n f}$ of the new program $P_3$, is constructed out
of that of $P_2$ by transforming clause (\ref{clause8.ex.P2})
as described in Approach 2:
\begin{equation}
\label{clause2.ex.P3}
\begin{minipage}{.53\textwidth}
\begin{verbatim}
sat_cnf( [Clause|Clauses], L ) :-
        sat_cl( Clause, L, -1 ),
        Lnew is L+1,
        catch( sat_cnf( Clauses, Lnew ),
               L,
               fail
              ).
\end{verbatim}
\end{minipage}
\end{equation}
So backjumping related to the variable with level $l$, implemented as 
$throw(l)$, arrives to an instance of clause (\ref{clause2.ex.P3})
 where %
$L$ is $l$.  The whole $catch(\ldots)$
fails, and the control backtracks to
the invocation of ${\it sat\_c l}$ that assigned the variable.
(An additional predicate ${\it b t id}$ was not needed, as
$L$ is the unique identifier.)

  Now program $P_3$ consists of clauses
(\ref{clause1.ex.P2}) -- (\ref{clause7.ex.P2}) and
(\ref{clause.throw.ex.P3}) -- (\ref{clause2.ex.P3}). 
To avoid leaving unnecessary backtrack points in some Prolog systems,
each group of clauses with ${\it var}$/1 and ${\it non var}$/1 
(clause (\ref{clause1.ex.P2}) with (\ref{clause2.ex.P2}), and 
(\ref{clause1aux.ex.P2}) with (\ref{clause2aux.ex.P2}) and
(\ref{clause3aux.ex.P2}))
may be replaced by a single
clause employing
$({\it var(V) \mathop\to \ldots{;}\ldots })$ and, in the second case
additionally $({\it H{<}L \mathop\to \ldots{;}\ldots })$.
To simplify a bit the initial queries, a top level predicate may be added,
defined by a clause \ \
{\small\verb|sat(Clauses)|\,\verb|:-|\,\verb|sat_cnf(Clauses,0).|}

\end{example}

\pagebreak[3]
\begin{example}\rm
\label{ex.binary}
    Here we transform $P_1$ from Ex.\,\ref{ex.approach2}  to a binary program and apply Approach 1.
    The binary program $P_{\mathrm b}$ is
    \[
   \begin{minipage}{.78\textwidth}  %
\begin{verbatim}
sat_b( [] ).
sat_b( [[Pol-Pol|_]|Clauses] ) :- sat_b( Clauses ).
sat_b( [[_|Pairs]|Clauses] ) :- sat_b( [Pairs|Clauses] ).
\end{verbatim}
   \end{minipage}
    \]
Note that %
in Ex.\,\ref{ex.approach2} the unprocessed part of the current clause was an
argument of ${\it sat\_cl}$, now it is the head of the %
 argument of ${\it sat\_b}$.
In what follows we do not explain some details which are as 
in the previous example.
As previously we introduce levels,
and represent a value of a variable by $(l, v)$, where $l$ is a 
level and $ v$ a logical value.  
As previously, we first transform  $P_{{\rm b}}$ into 
$P_{{\rm b}2}$ dealing with levels, and then add backjumping to $P_{{\rm b}2}$.
We add two arguments to ${\it sat\_b}$,
they are the same as the arguments added to  ${\it sat\_cl}$ in
Ex.\,\ref{ex.approach2}.  
The declarative semantics is similar, the first argument of 
${\it sat\_b}$ (in an answer of $P_{{\rm b}2}$) is
 as the first argument of ${\it sat\_c n f}$ in %
$P_2$. %
An invariant similar to that of Ex.\,\ref{ex.approach2} will be maintained
by the operational semantics.
Whenever ${\it sat\_b(cl s,l,h l)}$ is selected,
$l$ and $h l$ are numbers,
$l>h l$ and $l$ is greater than any number occurring in ${\it cl s}$.
List ${\it cl s}$ is a conjunction of clauses (possibly instantiated),
and its head, say $cl$, is the not yet processed fragment of the current
clause, say $cl_0$;
number $l$ is the current level, and
${\it h l}$ is the highest level of variables from the already processed part
of $cl_0$.
Now program $P_{{\rm b}2}$ is:
\[
\begin{minipage}[t]{.75\textwidth}
\begin{verbatim}
sat_b( [], _L, _HL ).
sat_b( [[Pol-V|_] | Clauses], L, _HL ) :- nonvar(V),  
        V=(_,Pol), Lnew is L+1,
        sat_b( Clauses, Lnew, -1 ).
sat_b( [[Pol-V|_] | Clauses], L, _HL ) :- var(V), 
        V=(L,Pol), Lnew is L+1,
        sat_b( Clauses, Lnew, -1 ).
sat_b( [[_-V|Pairs] | Clauses], L, HL ) :- 
        Lnew is L+1,
        new_highest( V, HL, HLnew ),
        sat_b( [Pairs | Clauses], Lnew, HLnew ).
\end{verbatim}
\end{minipage}
\makebox[0pt][l]{\hspace{2.8em}
    \begin{minipage}[t]{.05\textwidth}
        (\refstepcounter{equation}\theequation\label{clause1})%
    \label{program.exbinary2}%
    \label{clause1.ex.binary2}%
\\\\
        (\refstepcounter{equation}\theequation)%
    \label{clause2.ex.binary2}%
\\\\\\        (\refstepcounter{equation}\theequation)%
    \label{clause3.ex.binary2}%
\\\\\\ [1ex]        (\refstepcounter{equation}\theequation)
    \label{clause4.ex.binary2}
    \end{minipage}
}
\]
Procedure ${\it new\_highest}$/3 is the same as in the previous example.
Program $P_{{\rm b}2}$ with a query ${\it sat\_b}(sat,0,-1)$
checks satisfiability of the conjunction of clauses $sat$.

Now we add backjumping to $P_{{\rm b}2}$.  
Approach 1 is applicable, as each backjump target of interest consists of a
single atom ${\it sat\_b}(\ldots)$.
As previously, backjumping originates 
when an empty clause is encountered:
\begin{equation}
\label{clause5throw.ex.binary2}
\begin{minipage}{.75\textwidth}
\begin{verbatim}
sat_b( [[] | _Clauses], _L, HL ) :-  HL>=0, throw( HL ).
\end{verbatim}
\vspace{-1\abovedisplayskip}
\end{minipage}
\end{equation}
\commenta{
    Let us discuss backjump targets.
    Assume that the nodes of an LD-tree satisfy the invariant.
    Consider the descendants of a node $N = {\it sat\_b(cl s,l,h l)}$ 
     obtained by first resolving $N$ with clause (\ref{clause2.ex.binary2}) or
    (\ref{clause4.ex.binary2}).
    A ball thrown from such a descendant $N_2$ is not $l$.%
}% \commenta
Let us discuss backjump targets.  Speaking informally, 
we backjump to a place where a variable obtained a value; this happens only
in clause  (\ref{clause3.ex.binary2}).  
% So for the backjump target only this clause has to be modified.
So only this clause has to be modified to implement backjump target.
For a proof,
assume that the nodes of an LD-tree satisfy the invariant.
Consider the descendants of a node $N = {\it sat\_b(cl s,l,h l)}$ 
 obtained by first resolving $N$ with clause (\ref{clause2.ex.binary2}) or
(\ref{clause4.ex.binary2}).
A ball thrown from such a descendant $N_2$ is not $l$.%
\footnote{
  Consider a node $N = {\it sat\_b}(cl s,l,h l)$ and its closest descendant $N'$
   of the form ${\it sat\_b(\ldots)}$.
   So $N'={\it sat\_b(\ldots,l{+}1,\ldots)}$.
   If a number $i$ occurs in a node between $N$ and $N'$, or in $N'$,
    then $i=l{+}1$ or $i$ occurs in $N$.
   By induction, if $i$ occurs in a descendant of $N$ then $i$ occurs in $N$ or
   $i>l$. 
   Additionally, if $N'$ was obtained by first resolving $N$ with
   (\ref{clause2.ex.binary2}) or (\ref{clause4.ex.binary2}), then $N'$ does not
   contain $l$.  
   Thus no descendant of $N'$ contains $l$.
}  % \footnote
So (\ref{clause2.ex.binary2}), (\ref{clause4.ex.binary2}) transformed to the
form (\ref{eq.approach1.1}) will never catch a ball, and transforming them is unnecessary.

% So for the backjump target we need to modify only clause
% (\ref{clause3.ex.binary2}).  Following Section \ref{sec.approach1} we obtain:

% Clause (\ref{clause3.ex.binary2}) in the form  (\ref{eq.approach1.1}) 

Out of  (\ref{clause3.ex.binary2}) we obtain:
\begin{equation}
\label{clause3catch.ex.binary2}
\begin{minipage}{.75\textwidth}
\begin{verbatim}
sat_b( [[Pol-V|_] | Clauses], L, _HL ) :- 
        catch( ( var(V), V=(L,Pol), Lnew is L+1,
                 sat_b(Clauses, Lnew, -1)
               ),
               L,
               fail
             ).
\end{verbatim}
%         catch(
%              ( var(V), V=(L,Pol), Lnew is L+1,
%
%              fail
%              ).
\end{minipage}
\end{equation}
The final program $P_{{\rm b}3}$  consists of clauses
(\ref{clause1.ex.binary2}), (\ref{clause2.ex.binary2}), 
(\ref{clause3catch.ex.binary2}),
(\ref{clause4.ex.binary2}),  (\ref{clause5throw.ex.binary2}), 
and (\ref{clause1aux.ex.P2}) -- (\ref{clause3aux.ex.P2}).%

% \smallskip
  \vspace{0.5ex}
% \footnote
{
\small  
In (\ref{clause3catch.ex.binary2}), the first atom ${\it var}(V)$
  from the body of (\ref{clause3.ex.binary2})
  can be moved outside of ${\it catch}$,  %
  transforming the body of (\ref{clause3catch.ex.binary2}) to
  ${\it var(V), catch(\ldots)}$.
  (This is because ${\it var}(V)$ is deterministic and not involved in
  backjumping.)  Now, similarly as in the previous example, 
  some backtrack points may be avoided by replacing
  clauses (\ref{clause3catch.ex.binary2}) and (\ref{clause2.ex.binary2}) 
  by a single clause with the body
  {\tt%
    \renewcommand{\myunderline}{\symbol{95}}%
%  (var(V) -> V=(L,Pol) ; V=(\myunderline,Pol)),
 (var(V)->V=(L,Pol);V=(\myunderline,Pol)),
    Lnew is L+1,
%        sat\myunderline b( Clauses, Lnew, -1 ).
        sat\myunderline b(Clauses,Lnew,-1).
  } %\tt

}  % former \footnote
\smallskip

% \pagebreak %
%
Let us also mention that applying Approach 1a to $P_{{\rm b}2}$ results in
adding clause (\ref{clause5throw.ex.binary2}) and
replacing clauses 
    \ref{clause2.ex.binary2} --     \ref{clause4.ex.binary2} by
\[
\begin{minipage}[t]{.9\textwidth}
\small
\begin{verbatim}
sat_b( [[Pol-V|Pairs] | Clauses], L, HL ) :- 
   catch( (nonvar(V), V=(_,Pol), Lnew is L+1, sat_b(Clauses, Lnew, -1)
           ; throw(L) 
          ),
          L,
          catch( (var(V), V=(L,Pol), Lnew is L+1, sat_b(Clauses, Lnew, -1)
                  ; throw(L)
                 ),
                 L,
                 catch( (Lnew is L+1, new_highest(V, HL, HLnew),
                         sat_b([Pairs|Clauses], Lnew, HLnew)
                         ; throw(L)
                        ),
                        L,
                        fail
                      ) ) ).
\end{verbatim}
\end{minipage}
\]
\end{example}

\section
{Simulating backjumping by backtracking}

\noindent
This section complements the current paper by presenting an approach not
based on exception handling.
We discuss simulating backjumping by means of Prolog backtracking,
as suggested by
\citeN{DBLP:journals/tplp/Bruynooghe04}.
A backjump is initiated by a failure preceded 
by depositing in the Prolog
database an identifier of the backjump target.
At each backtracking step, the database is queried to check if 
backjumping is being performed and if its target is reached;
further backtracking is caused if necessary.
This is done by some extra code placed at the beginning of the body
of each clause involved in backjumping.
(In the presented example \cite{DBLP:journals/tplp/Bruynooghe04}, there is
only one such clause.) 

The paper by Bruynooghe is focused on a single example,
here we make it explicit how to implement the idea in a general case.
Let us introduce a predicate ${\it catch}$/1, to deal with backtracking that
simulates backjumping.
The role of
$catch(t)$ is to succeed immediately, unless during backjumping.  In the latter
case $catch(t)$ fails
if $t$ is not unifiable with (the identifier of) the backjumping target.
In this way the backjumping is continued.
Otherwise it removes the target from the database
and succeeds (instantiating $t$ in the obvious way).  This means completing
the backjump.

We can use\,  ${\it assert(target(t')), fail}$\, to cause a backjump, and define
\[\it 
 catch(Id)\; \gets\; target(\myunderscore)
 \begin{array}[t]{@{}c@{}ll}
  {}\to{}  & \it retract(target(Id))  & \mbox{\% within backjumping}
   \\
     ; & \it true     & \mbox{\% no backjumping}
  \end{array}
\]
Query
% {\tt retract(target($t$))}
${\it retract(target(t))}$ fails when $t$ is not unifiable
with the recorded target $t'$.
Otherwise it succeeds and removes the database
item, in this way indicating the end of the backjump.

To maintain such simulated backjumping, we convert
 each clause $p(\vec t)\gets B$ of the program  into
%%%%
 \vspace{-.4\medskipamount}
\begin{equation}
\label{clause.approach3}
{\it
p(\vec t)\gets b t id(\vec t,Id),catch(Id), B.  
}
\vspace{.5ex}
\end{equation}
where $b t id$/2 is as in Section \ref{sec.implementing}.
Let us present an informal explanation.
Note first that if the database is empty then the behaviour of
(\ref{clause.approach3}) does not differ from that of the original clause.
Assume now that backjumping has been initiated, so $target(t')$ is asserted and
backtracking has started.  At each backtrack point, a clause of the form 
(\ref{clause.approach3}) is involved. 
If ${\it catch}$ finds that the backjumping target is reached, $B$ is executed.
Otherwise  ${\it catch}$ fails, and the simulated backjump continues.

Often for many clauses of the program it is known that ${\it catch(Id)}$ in 
(\ref{clause.approach3}) will not catch any backjump.  In such case the
clause can be simplified to
\begin{equation}
\label{clause.approach3a}
{\it
 p(\vec t)\gets  %
\myprologneg(target(\myunderscore)), B.  
}
\end{equation}

Note that there are no restrictions 
in this approach 
on the origin or target of
backjumping, in contrast to those discussed in Section \ref{sec.implementing}.

\section{Final comments}
\label{sec.final}
\paragraph{Related work.}
For the approach of \citeN{DBLP:journals/tplp/Bruynooghe04}, see the previous
section.
\citeN{DBLP:journals/tplp/RobbinsKH21} 
do not present any general approach, but they show
a sophisticated example of 
using Prolog exception handling to implement backjumping.
(The main example is preceded by a simple introductory one.)
The program is a SAT solver with conflict-driven clause learning.
A learned clause determines the target of a backjump.
Note that the issue dealt with is not exactly backjumping, understood as in
Section \ref{sec.backjumping}.  In the SAT solver,
not only a fragment of the SLD-tree is to be skipped, but additionally the tree
has to be extended, as a new clause is added to the SAT problem.

The program employs Prolog coroutining in a fundamental way.
It uses exception handling also to deal with plain backtracking.
It keeps the learned clauses in the Prolog database, to preserve
them during backjumping.
The program is rather complicated;
it seems impossible to view it as some initial program with added
backjumping.  To understand it one has to reason about the details of the
operational semantics.  
This is not easy, due to sophisticated interplay of
  coroutining and exception handling.
(The involved semantic issues are discussed here in \ref{sec.semantics}.)

That paper  does not propose any general way of adding backjumping to logic
programs.  The difference between backjumping and Prolog exception handling
discussed here in Section \ref{sec.backjumping}  is not noticed.%
\footnote{
    This may be due to not facing the limitations
    pointed out here.
    Backjumping in the program of \citeN{DBLP:journals/tplp/RobbinsKH21}
    is similar to that of
    (\ref{program.approach1asmall})
    (Approach 1a for $n=2$),
     with ${\it throw(Id)}$  dropped 
     (as there is no standard backtracking), and 
     ${\it catch(B_2,Id,fail)}$ replaced by $B_2$
     (as there is no backjumping from $B_2$ with the current $Id$).
 %%%
} %  \footnote
We cannot agree with the claims ``backjumping is exception handling"
 and that ``{\tt catch} and {\tt throw} [provide]
exactly what is required for programming backjumping''
\cite[the title, and p.\,142-143]{DBLP:journals/tplp/RobbinsKH21}.

\paragraph{Conclusions.}
The subject of this paper is adding backjumping to logic programs.
We discussed
the differences between backjumping and Prolog
exception handling, and
showed that implementing the former by the latter is impossible in a general
case (Section \ref{sec.backjumping-exceptions}).
We proposed two approaches to such implementation.
The first approach imposes certain restrictions on where backjumping can be
started.  The second one  -- on the target of backjumping.
The restrictions seem not severe.
The first approach is applicable, among others, to binary programs with
arbitrary backjumping.  
For the second approach, 
the presented example shows that sometimes the difference between the
required and the actual target
may be unimportant.
As every program can be transformed to a binary one
\cite{DBLP:books/mk/minker88/Maher88,DBLP:conf/plilp/TarauB90},
the first approach is indirectly applicable to all cases.
%

%
%
%
%
%%%%
% \vspace{-1.5ex}
\paragraph{\bf Acknowledgement.}
Thanks are due to the anonymous reviewers for their stimulating comments.
Jan Wielemaker, David Geleßus and Ed Robbins commented on the content of the
Appendix.
\commenta{...
  and pointing a few ugly small errors
  }

%%%
% \vspace{-1.5ex}
\subparagraph{\bf Competing interests.}
The author declares none.

%%%
% \vspace{-.5ex}
\appendix
\section{Exception handling in the presence of coroutining}
\label{sec.semantics}

In this paper we considered Prolog without coroutining.
Coroutining, also known as delays,  is a particular way of
 modifying the Prolog selection rule.  The first atom of 
% the query may be not selected
the query may not be selected (we say that it is delayed, or blocked).
The Prolog built-ins to deal with delays are {\tt when}/2, {\tt freeze}/2
(and {\tt block} declarations of SICStus).

The behaviour of exception handling combined with delays seems far from
obvious.   
Programs using both these features, like that of
\citeN{DBLP:journals/tplp/RobbinsKH21}, may be difficult to understand.
Explanations are difficult to find.  Delays are outside of the
scope of the Prolog standard.
So it should be useful to provide a formal description.
For this we first describe the Prolog coroutining, in terms of
SLD-resolution. 

%%%
% \vspace{-.5ex}
\paragraph{Semantics of coroutining.}
This description is restricted to the built-in {\tt when}/2.  The other
constructs modifying the selection rule can be expressed in terms of
{\tt when}/2.

A query ${\it when(C,Q)}$ blocks the query $Q$ until the condition $C$ is
true.  
An example condition is ${\it non var}(X)$, it blocks Q until $X$ is bound to
a non-variable term.
For the possible form of the condition, we refer the reader to Prolog manuals. 

The nodes of SLD-trees are queries; %
queries are, as usual, sequences of 
atoms.  However in addition to atoms of the underlying
logical language, atoms of the form  ${\it when(C,Q)}$ can be used.
(Note that this is a recursive definition, as $Q$ is a query.)
At the beginning of a query there may appear some ${\it when}$-atoms that
are known to be blocked.  We separate them by symbol $\&$ from the rest of
the query, 
and we call them the {\em blocked part} (or delayed part) of the query.
We may skip  $\&$ when the blocked part is empty.
The rest of the query is its  {\em active part}.
An invariant will be maintained that in a query
${\it when(C_1,Q_1),\ldots,when(C_n,Q_n)\&Q}$
each condition $C_i$ ($1\leq i\leq n$) is not satisfied.

The selection rule of SLD-resolution selects in each query
the first atom after $\&$, i.e. the first atom of the active part of a query.
The atom will be called the {\em selected atom} of the query. 
Now we are ready to describe a resolution step.

\pagebreak[3]
\begin{definition}
\label{def.SLD}
A {\em successor} $\QQ'$ of a query $\QQ=D\&A,Q$, where $A$ is the selected atom,
is obtained by an extension of an SLD-resolution step as follows.
\pagebreak[3]
\begin{enumerate}
\setlength{\leftmargini}{.7\leftmargini}
\item 
If $A$ is ${\it when(C,Q')}$ then
  \begin{enumerate}
  \item 
  if condition $C$ is satisfied then $\QQ'$ is $D\&Q',Q$,
  \item
  otherwise $\QQ'$ is $D,A\&Q$,
  \end{enumerate}
\pagebreak[3]
\item 
If $A$ is not a ${\it when}$-atom then
  \begin{enumerate}
  \item 
  a standard SLD-resolution step is performed (unification of  $A$ with a
  clause head, replacing $A$ with the clause body $B$, applying the 
    mgu to the resulting query).  This produces $Q_1=(D\&B,Q)\theta$.
  \item
  Let $D_{\rm u n bl}$ be those ${\it when}$-atoms from $D\theta$  whose
  conditions are true, and $D'$ be those whose conditions are false.
  Now $\QQ'$ is\,  $D'\& D_{\rm u n bl},B\theta,Q\theta$.
  {\sloppy\par}
  \end{enumerate}
\end{enumerate}
\end{definition}

In this way we defined the notions of SLD-derivation and SLD-tree for Prolog
in the presence of coroutining. 
Note that the order of elements of 
$D_{\rm u n bl}$ is not specified. 
So we do not specify
the order of unblocking when several queries are unblocked in the same
resolution step.  Such details are not %
described by the documentation of
Prolog systems.
(SWI-Prolog seems to preserve 
in $D'$ and in $D_{\rm u n bl}$ the order of atoms from $D$.) 

Let us define (similarly as in Section \ref{sec.introduction})
the {\em execution}
of a query $\QQ=D\&A,Q$ (in an SLD-derivation \DD) to be the part $\DD'$ of \DD
consisting of \QQ 
and the following queries of the form $D'\&Q',Q\theta$, up to the first query
of the form $D'\&Q\theta$, if such query exists. 
If the last query of $\DD'$ is  $D'\&Q\theta$ 
(i.e.\ $\DD'$ is {\em  successful}\/)
and $\theta$ is the composition of the mgu's used in $\DD'$,
then $A\theta$ will be called a {\em pseudo-answer} for $A$ (in $\DD'$).
The Prolog debugger displays such pseudo-answer at the Exit port corresponding
to the Call port for A.  
% If \QQ is the initial query (and $D,Q$ are empty),
If  $D,Q$ are empty and $\QQ=A$ is the initial query,
then Prolog displays (in the standard encoded form)
a pseudo-answer $A\theta$, augmented with $D'$.

% (as usually, $A\theta$ is encoded as a set of equations).
% (As usually, $A\theta$ is encoded as a set of equations, corresponding to a
% relevant subset of $\theta$).

\paragraph{Semantics of exception handling.}
Now we can specify the semantics of Prolog exception handling in the presence
of delays, as described in Def. \ref{def.SLD}.
The following modifications to the description of Section \ref{sec.exceptions}
are sufficient.
\begin{itemize}
\item
We consider %
  $A_c={\it catch(Q,s,Handler)}$  selected in a node $D\&A_c,N'$;
  its child is $D\&Q,N'$.
\item
  The node raising the exception is  $N_t=D_t\&{\it throw(t),N_t'}$.
\item
  The search is for a node %
  $N_c=D'\&{\it catch(Q',s',Handler')},N_c'$.

\item  
  Condition (b) is modified to

  \qquad
\begin{oldtabular}{l}
  (b) no node between $N_c$ and $N_t$ (including  $N_t$)  is of the form
  $D''\&N_c'\theta$. 

\end{oldtabular}

\end{itemize}

\bibliographystyle{acmtrans}
\bibliography{bibshorter,backjumping.arxiv,bibmagic}

\end{document}